\documentclass{article}
\usepackage{waspaa21,amsmath,graphicx,url,times}
\usepackage{color}

\usepackage[linesnumbered,boxed,titlenumbered,ruled,nofillcomment]{algorithm2e}
\usepackage[mathscr]{eucal}
\usepackage{bm}
\usepackage{bbm}
\usepackage{amsthm}
\usepackage{amssymb}
\usepackage{mathtools} \mathtoolsset{showonlyrefs=false} \MakeRobust{\eqref}
\usepackage{graphicx}
\usepackage[hidelinks]{hyperref}
\usepackage[dvipsnames]{xcolor}
\usepackage{enumerate}
\usepackage[utf8]{inputenc}
\usepackage{booktabs}

\SetInd{0.5em}{0.5em}

\hypersetup{
	colorlinks=false,  
	citecolor=MidnightBlue,
	urlcolor=MidnightBlue,
	linkcolor=MidnightBlue,
}

\SetCommentSty{mycommfont}

\theoremstyle{plain}

\theoremstyle{definition}

\newcommand{\F}{\textup{F}}

\newcommand{\zerobf}[0]{\bm{0}}

\newcommand{\Gammabf}[0]{\bm{\Gamma}}

\newcommand{\Lambdabf}[0]{\bm{\Lambda}}
\newcommand{\rhobf}[0]{\bm{\rho}}
\newcommand{\mubf}[0]{\bm{\mu}}
\newcommand{\varepsilonbf}[0]{\bm{\varepsilon}}
\newcommand{\Abf}[0]{\bm{A}}

\newcommand{\Ibf}[0]{\bm{I}}

\newcommand{\Mbf}[0]{\bm{M}}

\newcommand{\Qbf}[0]{\bm{Q}}
\newcommand{\Rbf}[0]{\bm{R}}

\newcommand{\ebf}[0]{\bm{e}}

\newcommand{\xbf}[0]{\bm{x}}
\newcommand{\ybf}[0]{\bm{y}}

\newcommand\Eb{\mathbb{E}}

\newcommand\Rb{\mathbb{R}}
\newcommand\Sb{\mathbb{S}}

\DeclareMathOperator*{\argmin}{arg\,min}

\newcommand{\defeq}{\stackrel{\text{\tiny \textnormal{def}}}{=}}  

\newcommand{\diag}[0]{\operatorname{diag}}

\newcommand{\proj}[0]{\operatorname{proj}}
\newcommand{\EIR}[0]{\operatorname{EIR}}

\newcommand{\sym}[0]{\operatorname{sym}}

\usepackage{tikz}
\newcommand\copyrighttext{%
	\scriptsize 
	\copyright 2021 IEEE. Personal use of this material is permitted. Permission from IEEE must be obtained for all other uses, in any current or future media, including reprinting/republishing this material for advertising or promotional purposes, creating new collective works, for resale or redistribution to servers or lists, or reuse of any copyrighted component of this work in other works. 
	The published version of the paper is: 
	O.\ A.\ Malik, V.\ V.\ Narumanchi, S.\ Becker and T.\ W.\ Murray, ``Superresolution Photoacoustic Tomography Using Random Speckle Illumination and Second Order Moments,'' 2021 IEEE Workshop on Applications of Signal Processing to Audio and Acoustics (WASPAA), 2021, pp.\ 141--145, \url{https://doi.org/10.1109/WASPAA52581.2021.9632758}.
	}
\newcommand\mycopyrightnotice{%
	\begin{tikzpicture}[remember picture,overlay]
		\node[anchor=south,yshift=20pt] at (current page.south) {\fbox{\parbox{\dimexpr\textwidth-\fboxsep-\fboxrule\relax}{\copyrighttext}}};
	\end{tikzpicture}%
}

\title{Superresolution photoacoustic tomography using random\\speckle illumination and second order moments}

\name{Osman Asif Malik,$^{1}$\sthanks{This material is based upon work supported by the National Science Foundation under Grant No.\ 1810314.}
      Venkatalakshmi Vyjayanthi Narumanchi,$^{2}$$^*$
      Stephen Becker,$^{1}$$^*$
      Todd W.\ Murray$^{3}$$^*$
  }
\address{
$^1$ University of Colorado Boulder, Dept.\ of Applied Mathematics, Boulder, CO, USA\\ 
osman.malik@colorado.edu, stephen.becker@colorado.edu\\
$^2$ University of Colorado Boulder, Dept.\ of Electrical Engineering, Boulder, CO, USA\\
vyjayanthi.narumanchivenkatalakshmi@colorado.edu\\ 
$^3$ University of Colorado Boulder, Paul M.\ Rady Dept.\ of Mechanical Engineering, Boulder, CO, USA\\
todd.murray@colorado.edu
}

\begin{document}

\ninept
\maketitle

\begin{sloppy}

\begin{abstract}
Idier et al.\ [IEEE Trans.\ Comput.\ Imaging 4(1), 2018] propose a method which achieves superresolution in the microscopy setting by leveraging random speckle illumination and knowledge about statistical second order moments for the illumination patterns and model noise.
This is achieved without any assumptions on the sparsity of the imaged object.
In this paper, we show that their technique can be extended to photoacoustic tomography.
We propose a simple algorithm for doing the reconstruction which only requires a small number of linear algebra steps.
It is therefore much faster than the iterative method used by Idier et al.
We also propose a new representation of the imaged object based on Dirac delta expansion functions.
\end{abstract}

\begin{keywords}
Photoacoustic tomography (PAT), medical imaging, superresolution imaging, blind speckle illumination, second order statistics
\end{keywords}

\section{Introduction} \label{sec:intro}

Photoacoustic tomography (PAT) is a technique which detects optical contrast acoustically. 
As photons travel through an absorbing medium like tissue, most, if not all the light is converted into heat. 
This causes a temperature increase which induces a pressure increase through thermo-elastic expansion. 
This pressure then dissipates through the tissue as a wideband acoustic signal. 
An ultrasound transducer array records these signals which can be used to form an image. 
This method has found several uses, as one of the key advantages of PAT over purely optical methods is the increase 
in imaging depth within tissue since acoustic signals scatter less than light.
A review of PAT imaging methods can be found in \cite{yao2013}. 
In the particular imaging method that is the focus of this paper, the illumination is generally a diffuse beam, and a single ultrasound transducer or a transducer array is scanned to obtain an image. 
This is illustrated in Fig.\ \ref{fig:pat-illustration}.
The resolution is limited by the focal volume of the transducer which is at least two orders of magnitude greater than the wavelength of light. 
This resolution limit decreases as the imaging depth increases. 
For this reason, 
there is a need for techniques that can improve the imaging capability at increased depths.

\begin{figure}[ht!]
	\centering  
	\includegraphics[width=1\columnwidth]{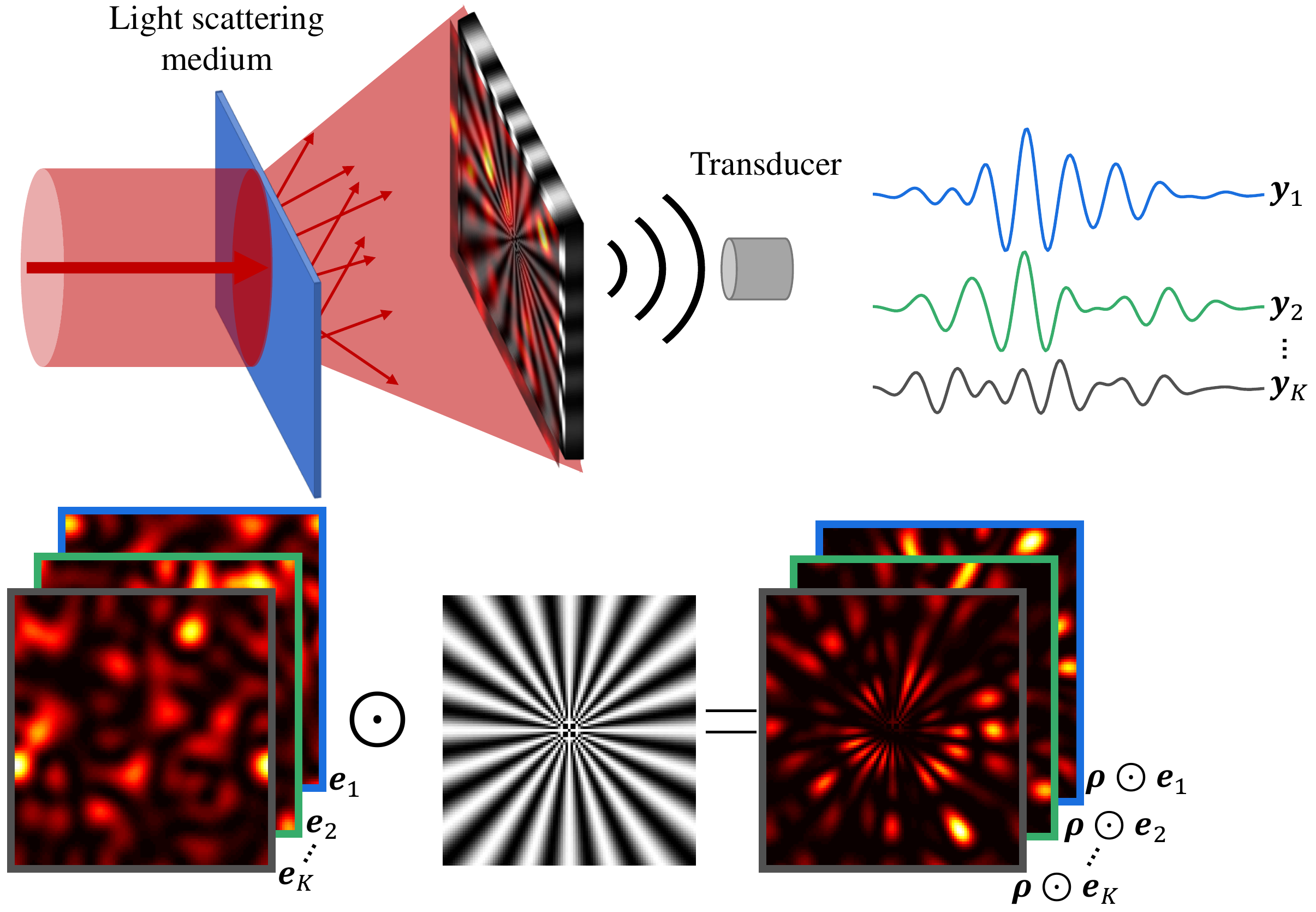}
	\caption{The PAT setting with different speckle illuminations.}
	\label{fig:pat-illustration}
\end{figure}

When light enters a scattering medium like tissue, the photons are scattered from their initial paths and form random speckle patterns \cite{sebbah2001waves}.
By leveraging the recordings generated by many such random illumination patterns, the resolution can be improved in both microscopy and PAT \cite{chaigne2016, hojman2017a, murray2017, mudry2012b, min2013a}.
Our paper is inspired by the work done by Idier et al.\ \cite{idier2018} in the microscopy setting.
They demonstrate that random speckle illumination and the second order moments of the speckle and noise can be used to improve resolution in certain regimes---without any assumptions on the sparsity of the imaged object.
In particular, they show that decreasing the size of the speckle patterns can lead to improved resolution.

We make the following contributions in this paper:
\begin{itemize}
	\item We provide evidence that the ideas in \cite{idier2018} can be extended to PAT, and that random speckle illumination combined with second order moment information can lead to enhanced resolution in PAT.
	
	\item We represent the imaged object using Dirac delta expansion functions.
	This representation allows us to use a coarser temporal discretization than e.g.\ spherical expansion functions, leading to faster computation and less memory usage.
	\mycopyrightnotice
	\item We propose a simple algorithm for recovering the object from the empirical transducer recording covariance matrix. 
	Unlike the iterative algorithm in \cite{idier2018}, our method requires a small number of simple linear algebra steps and therefore runs much faster (minutes instead of hours). 
\end{itemize}

\section{Related work} \label{sec:related-work}

In numerous recent works in PAT, the conventional resolution limits are surpassed via
methods using speckle illumination \cite{gateau2013, chaigne2016, hojman2017a, murray2017, liu2019a}, SOFI \cite{dertinger2009} inspired fluctuation imaging \cite{chaigne2017, vilov2020, vilov2020a}, and localization-based methods \cite{vilov2017, dean-ben2018}. 
The class of methods \cite{gateau2013, chaigne2016, hojman2017a, murray2017, liu2019a} that has taken inspiration from blind structured illumination microscopy \cite{mudry2012b, min2013a} has achieved a resolution improvement by a factor of two. 
These techniques illuminate the sample with unknown speckle patterns instead of a uniform beam. 
This has the effect of frequency shifting the acoustic signals into a frequency band which is detectable by the transducer. 
Thus, the grain size of the speckle and the transducer properties have a significant impact on the achievable resolution. 
Previous approaches \cite{chaigne2016, murray2017} 
estimate the object as the solution to a regularized optimization problem, solved via iterative methods. 
The regularizer is chosen to take advantage of the fact that the object is usually sparse.
There has been other work \cite{yeh2017a} that also uses second order speckle information. 
In this purely optical method, the illumination patterns are initially estimated after calculating a widefield low-resolution image. 
The covariance of these recovered patterns are then used to reconstruct a covariance image which estimates the object. 

Many optical imaging techniques extend to PAT but there are key differences, for example 
PAT has a non-uniform point response unlike in optical microscopy, hence the resolution is different along the axial and lateral directions. In addition to this, the point response is uniform only within a small region of the imaging aperture and it is desirable to have a general forward model that takes this into account,
in contrast to optical microscopy where a point spread function can be defined for the entire field of view.

\section{Reconstruction via second order moments} \label{sec:method}

Let $\rhobf \in \Rb^N$ be a vector representing the object we are trying to reconstruct.
For example, on a two-dimensional grid of size $n \times n$ we have $N = n^2$.
Moreover, let $\ebf \in \Rb^N$ be a vector describing the speckle pattern illuminating the object.
In the microscopy setting in \cite{idier2018}, the recorded image $\ybf$ can be modeled as
\begin{equation} \label{eq:idier-model}
	\ybf = h * (\rhobf \odot \ebf) + \varepsilonbf,
\end{equation}
where $h$ is a point spread function (PSF), $\varepsilonbf$ is noise, $*$ denotes convolution, and $\odot$ denotes Hadamard (pointwise) product.

In this paper, we consider a PAT setup with $M$ transducers, each recording a time series consisting of $T$ time points for each of $K$ different random speckle patterns.
The recorded signal for a given speckle pattern $\ebf$ may now be modeled as
\begin{equation} \label{eq:model}
	\ybf = \Abf (\rhobf \odot \ebf) + \varepsilonbf = \Abf \Rbf \ebf + \varepsilonbf,
\end{equation}
where $\Rbf \defeq \diag(\rhobf)$, $\ybf \in \Rb^{TM}$ contains the length $T$ time series recording for each transducer concatenated into a single vector, $\varepsilonbf \in \Rb^{TM}$ is a random noise vector, and $\Abf \in \Rb^{TM \times N}$ is a linear operator.
The form of $\Abf$ depends on how the imaged object is represented.
$\Abf$ can either be derived analytically from the photoacoustic wave equation and properties of the transducers \cite{wang2008, wang2011} or empirically by observing the recorded signal from a known point absorber \cite{egolf2018, vilov2020a}.
We use a variant of the forward operator derived in \cite{wang2011} which we describe in Section~\ref{sec:forward-model}.
However, our reconstruction method should work well with any reasonable forward model.

The goal of the reconstruction problem is to recover $\rhobf$. 
Let $\ybf^{(1)}, \ldots, \ybf^{(K)}$ denote recorded signals corresponding to $K$ different speckle patterns.
A standard assumption is that the speckle pattern intensity, on average, is the same for each point of the object and that the noise is centered around zero.
Mathematically, these assumptions can be written as $\Eb[e_n] = \mu$ for each $n \in \{1,\ldots,N\}$ where $\mu$ is some fixed number, and $\Eb[\varepsilonbf] = \zerobf$.
Under these assumptions, $\bar{\ybf} \defeq K^{-1} \sum_{k=1}^K \ybf^{(k)} \approx \Abf \rhobf$ if the number of speckles $K$ is sufficiently large.
One may then estimate $\rhobf$ by solving the following least squares problem:
\begin{equation} \label{eq:rho-hat-1}
	\hat{\rhobf}_1 = \argmin_{\rhobf \in \Rb^{N}} \| \Abf \rhobf - \bar{\ybf} \|_2 + \lambda \|\rhobf\|_2.
\end{equation} 
The subscript on $\hat{\rhobf}_1$ indicates that this estimate uses assumptions on the mean, or first order moment, of the speckles and noise. 
Similar assumptions are made in e.g.\ \cite{mudry2012b, murray2017}.
The added Tikhonov regularizer is necessary since the inversion is ill-posed.

Similarly to \cite{idier2018}, we make the the additional assumption that we know the speckle covariance matrix $\Gammabf_{\ebf} \defeq \Eb[(\ebf - \mubf_{\ebf}) (\ebf - \mubf_{\ebf})^\top]$, where $\mubf_{\ebf} \defeq \Eb[\ebf]$, and the noise covariance matrix $\Gammabf_{\varepsilonbf} \defeq \Eb[\varepsilonbf \varepsilonbf^\top]$ (recall that $\Eb[\varepsilonbf] = \zerobf$).
This amounts to an assumption on the \emph{second order moments}.
Additionally, we assume that the random speckle pattern $\ebf$ and the noise $\varepsilonbf$ are independent. 
It is then easy to show that the signal covariance matrix $\Gammabf_{\ybf}$ satisfies
\begin{equation} \label{eq:signal-covariance}
	\Gammabf_{\ybf} \defeq \Eb[(\ybf - \mubf_{\ybf}) (\ybf - \mubf_{\ybf})^\top] = \Abf \Rbf \Gammabf_{\ebf} \Rbf \Abf^\top + \Gammabf_{\varepsilonbf},
\end{equation}
where $\mubf_{\ybf} \defeq \Eb[\ybf]$.
We compute the empirical covariance matrix $\hat{\Gammabf}_{\ybf}$ via
\begin{equation}
	\hat{\Gammabf}_{\ybf} = \frac{1}{K} \sum_{k=1}^K \ybf^{(k)} \ybf^{(k) \top} - \hat{\mubf}_{\ybf} \hat{\mubf}_{\ybf}^\top, \;\;\;\; \hat{\mubf}_{\ybf} = \frac{1}{K} \sum_{k=1}^K \ybf^{(k)}. 
\end{equation}
After replacing the unknown $\Gammabf_{\ybf}$ in \eqref{eq:signal-covariance} with $\hat{\Gammabf}_{\ybf}$, we ``solve'' that equation for $\Rbf$ which contains the sought object $\rhobf$ on the diagonal.
Recovering $\Rbf$ from \eqref{eq:signal-covariance} is a nontrivial problem.
Idier et al.\ \cite{idier2018} propose using an iterative nonlinear conjugate gradient (CG) method to do this.
It requires inverting an estimate of $\Gammabf_{\ybf}$ which costs $O(N^3)$ \emph{per iteration}.
Nonlinear CG methods usually require hundreds of iterations, which makes the method very expensive.
In Section~\ref{sec:algorithm}, we provide details on our simple noniterative method for estimating $\Rbf$ from \eqref{eq:signal-covariance} which costs $O(N^3)$ \emph{in total}.
It only requires basic linear algebra computations which are easy to implement.

\subsection{The forward operator model} \label{sec:forward-model}

For the forward operator $\Abf$, we use a variant of the discrete-to-discrete operator proposed in \cite{wang2011} which incorporates a model for the acousto-electric impulse response (EIR) of ultrasound transducers. 
We use the EIR model in equation (6) of \cite{liu2012}.
Applying $\Abf$ na\"{i}vely to a vector costs $O(TMN)$ operations.
However, $\Abf$ can be split into two parts, $\Abf = \Abf_\text{EIR} \Abf_0$, where $\Abf_0$ gives the transducer recordings without the EIR and $\Abf_\text{EIR}$ then applies a convolution with the EIR \cite{wang2011}.
The benefit of this approach is that $\Abf_0$ usually is sparse and $\Abf_\text{EIR}$ can be applied implicitly in time $O(TM \log (TM))$ by using the FFT and the convolution theorem.

Our operator differs from that in \cite{wang2011} in the choice of expansion functions used to represent the object.
We may represent an object $f : \Rb^3 \rightarrow \Rb$ on grid points $\{\xbf^{(n)}\}_{n=1}^N \subset \Rb^3$ via
\begin{equation} \label{eq:f-approx}
	f(\xbf) \approx \hat{f}(\xbf) \defeq \sum_{n = 1}^N \rho_n \phi_n(\xbf),
\end{equation}
where $\{\phi_n\}_{n=1}^N$ is a family of expansion functions and $\rhobf = (\rho_n)_{n=1}^N$ is a vector of coefficients.
Wang et al.\ \cite{wang2011} choose each $\phi_n$ to be a spherical expansion function centered at $\xbf^{(n)}$.
We found that this choice works poorly when $\Abf$ is split up into two separate operators $\Abf = \Abf_\text{EIR} \Abf_0$.
The reason is that the ``N'' shaped signal that results after computing $\Abf_0 \rhobf$ require a very fine temporal grid (i.e., large $T$) for accurate representation. 
To address this, we instead use expansion functions $\phi_n(\xbf) \defeq \delta(\xbf - \xbf^{(n)})$.
Define 
\begin{equation} \label{eq:photoacoustics-model}
	s(\xbf, t) \defeq \frac{\beta}{4 \pi C_p} \sum_{n=1}^N \frac{\rho_n}{\|\xbf - \xbf^{(n)}\|} \delta \Big(t - \frac{\|\xbf - \xbf^{(n)}\|}{c_0}\Big),
\end{equation} 
where $\beta$ is the thermal coefficient of volume expansion, $C_p$ is the specific heat capacity of the medium at constant pressure, and $c_0$ is the speed of sound in the object and background medium.
The signal generated at position $\xbf$ by the approximation $\hat{f}$ in \eqref{eq:f-approx} at time $t$ when $\phi_n(\xbf) = \delta(\xbf - \xbf^{(n)})$ is then $\partial s(\xbf,t)/\partial t$.
Convolving with the EIR and using properties of the Dirac delta function, we get $\EIR * (\partial s /\partial t) = \EIR' * s$, where $\EIR' \defeq d \EIR / dt$.
We therefore split our forward operator into two parts $\Abf = \Abf_{\EIR'} \Abf_s$ where $\Abf_s$ transforms a discretized object $\rhobf$ to a discretized signal and $\Abf_{\EIR'}$ applies convolution with $\EIR'$. 
$\Abf_s$ is sparse and $\Abf_{\EIR'}$ can be applied implicitly via the  FFT, so $\Abf \rhobf$ can be computed efficiently.
Moreover, the representation $\Abf_s \rhobf$ performs well even on relatively coarse temporal grids.

\subsection{Reconstruction algorithm} \label{sec:algorithm}

Our reconstruction method is presented in Alg.~\ref{alg:recovery}.
After subtracting the noise covariance on line~\ref{line:sub-mean} and solving the systems on lines~\ref{line:solve-1} and \ref{line:solve-2}, $\Mbf_3$ approximates $\Rbf \Gammabf_{\ebf} \Rbf$.
After multiplying $\Mbf_3$ on each side by $\sqrt{\Gammabf_{\ebf}}$ on line~\ref{line:add-sqrt} and subsequently taking the square root in line~\ref{line:take-sqrt} (we discuss the symmetrization and projection steps below), $\Mbf_5$ approximates $\sqrt{\Gammabf_{\ebf}} \Rbf \sqrt{\Gammabf_{\ebf}}$.
After the two solves on lines~\ref{line:solve-3} and \ref{line:solve-4}, $\hat{\Rbf}$ approximates $\Rbf$.
Finally, on line~\ref{line:diag} the diagonal $\hat{\rhobf}$ estimating $\rhobf$ is extracted.

\begin{algorithm}[ht!] \label{alg:recovery}
	\DontPrintSemicolon 
	\SetAlgoNoLine
	\KwIn{Estimate $\hat{\Gammabf}_{\ybf}$; known $\Gammabf_{\ebf}$, $\Gammabf_{\varepsilonbf}$, $\Abf$; constants $\lambda_1, \lambda_2$}
	\KwOut{Object estimate $\hat{\rhobf}$}
	$\Mbf_1 = \hat{\Gammabf}_{\ybf} - \Gammabf_{\varepsilonbf}$\label{line:sub-mean}\;
	$\Mbf_2 = \argmin_{\Mbf} \| \Abf \Mbf - \Mbf_1 \|_\F^2 + \lambda_1 \|\Mbf\|_\F^2$\label{line:solve-1}\;
	$\Mbf_3 = \argmin_{\Mbf} \| \Mbf \Abf^\top - \Mbf_2 \|_\F^2 + \lambda_1 \|\Mbf\|_\F^2$\label{line:solve-2}\;
	$\Mbf_4 = \sqrt{\Gammabf_{\ebf}} \Mbf_3 \sqrt{\Gammabf_{\ebf}}$\label{line:add-sqrt}\;
	$\Mbf_5 = \sqrt{\proj_{\Sb_+^N} (\sym(\Mbf_4))}$\label{line:take-sqrt} \tcp*{Compute via \eqref{eq:projection-computation}}
	$\Mbf_6 = \argmin_{\Mbf} \| \sqrt{\Gammabf_{\ebf}} \Mbf - \Mbf_5 \|_\F^2 + \lambda_2 \| \Mbf \|_\F^2$\label{line:solve-3}\;
	$\hat{\Rbf} = \argmin_{\Mbf} \| \Mbf \sqrt{\Gammabf_{\ebf}} - \Mbf_6 \|_\F^2 + \lambda_2 \| \Mbf \|_\F^2$\label{line:solve-4}\;
	Set $\hat{\rhobf}$ to diagonal of $\hat{\Rbf}$ \label{line:diag}\;
	\Return{$\hat{\rhobf}$}\;
	\caption{Efficient reconstruction of $\hat{\rhobf}$ from \eqref{eq:signal-covariance}}
\end{algorithm}

Idier et al.\ \cite{idier2018} point out that since $\hat{\Gammabf}_{\ybf}$ is an empirical covariance matrix, it may not be positive semidefinite. 
Consequently, the matrix $\Mbf_4$ may not be positive semidefinite and its square root may not exist.
Idier et al.\ therefore propose using a Kullback--Leibler divergence based dissimilarity measure between the empirical and true distributions, and then find an estimate $\hat{\rhobf}$ via a nonlinear CG method.  
Additionally, due to the regularizers on lines \ref{line:solve-1} and \ref{line:solve-2}, $\Mbf_4$ may not be exactly symmetric.
We propose a very simple solution to address these challenges:
We symmetrize and then project $\Mbf_4$ onto the set of positive semidefinite $N \times N$ matrices before taking the square root on line~\ref{line:take-sqrt}.
The symmetrization can be done via $\sym(\Mbf_4) = (\Mbf_4 + \Mbf_4^\top)/2$.
Let $\Sb^N$ and $\Sb_+^N$ denote the symmetric and positive semidefinite matrices of size $N \times N$, respectively. 
The projection operator $\proj_{\Sb_+^N} : \Sb^N \rightarrow \Sb_+^N$ is defined as
\begin{equation} \label{eq:projection}
	\proj_{\Sb_+^N}(\Mbf) \defeq \min_{\Mbf' \in \Sb_+^N} \|\Mbf' - \Mbf\|_\F.
\end{equation}
This projection is easy to compute via
\begin{equation} \label{eq:projection-computation}
	\proj_{\Sb_+^N}(\Mbf) = \Qbf \max(\Lambdabf, 0) \Qbf^\top,
\end{equation}
where $\Mbf = \Qbf \Lambdabf \Qbf^\top$ is the eigendecomposition of $\Mbf$, and the $\max(\cdot, 0)$ operator is applied elementwise.
The projection is the same if spectral norm is used instead of Frobenius norm in \eqref{eq:projection}; see Section~8.1.1 of \cite{boyd2004} for details.
The matrix $\proj_{\Sb_+^N}(\sym(\Mbf_4))$ is positive semidefinite and therefore guaranteed to have a square root; see Theorem~7.2.6 in \cite{horn2012} for details.
In practice, we find that the square root of $\Mbf_4$ usually exists, in which case the symmetrization and projection steps can be skipped.

We found that the Tikhonov regularization in lines~\ref{line:solve-1}, \ref{line:solve-2}, \ref{line:solve-3} and \ref{line:solve-4} of Alg.\ \ref{alg:recovery} with a careful choice of $\lambda_1$ and $\lambda_2$ is essential for the reconstruction. 
The regularization terms can easily be incorporated into the design matrix. 
For example, the problem in line~\ref{line:solve-1} can be written as
\begin{equation} \label{eq:large-system}
	\Mbf_2 = \argmin_{\Mbf} \left\| \begin{bmatrix}\Abf \\ \sqrt{\lambda} \Ibf \end{bmatrix} \Mbf - \begin{bmatrix} \Mbf_1 \\ \zerobf \end{bmatrix} \right\|_\F^2.
\end{equation}
If the problem in line~\ref{line:solve-2} is transposed and rewritten in a similar fashion, it will have the same design matrix.
The leading order cost of solving these problems is decomposing the design matrix (e.g.\ via the QR decomposition; see Section~5.3.3 of \cite{golub2013} for details), and this therefore only has to be done once for both lines.
In fact, since $\Abf$ remains fixed for a certain imaging setup, the decomposition only needs to be computed once for that setup.
Similar cost savings are possible for the lines~\ref{line:solve-3} and \ref{line:solve-4}.
The leading order cost of our algorithm is decomposing the design matrix in \eqref{eq:large-system}, which costs $O(\max(TM,N) N^2)$.
If this has been done ahead of time for the particular imaging setup, the leading order cost is reduced to $O(N^3)$.

\section{Experiments} \label{sec:experiments}

We run simulation experiments in Matlab in which 
we compare our method to the first order reconstruction estimate $\hat{\rhobf}_1$ in \eqref{eq:rho-hat-1} and to time reversal image reconstruction in k-Wave \cite{treeby2010} based on the average signal $\bar{\ybf}$.
The object we try to recover is the star-shaped object of size 160 \textmu m by 160 \textmu m shown in Fig.~\ref{fig:square-array}~(a).
In order to avoid inverse crime, we use different grids to represent the object when we generate the data and when we do the reconstruction. 
For data generation we use a 101 by 101 grid ($N=101^2$) and for reconstruction we use an 81 by 81 grid ($N=81^2$).
We use $M=64$ transducers arranged in two different geometries shown in Fig.~\ref{fig:transducer-geometries}.
In the first geometry, the transducers are arranged into a square array positioned a distance 30 \textmu m above the object.
In the second geometry, the transducers are positioned in a circle of radius 160 \textmu m around the object and in the same plane as the object.
The transducers have a center frequency $f_0 = \text{50 MHz}$ and full width at half-maximum $\text{FWHM} = \text{25 MHz}$.
Fig.~\ref{fig:transducer-geometries} shows examples of transducer recordings.
The recordings are 199 ns long and discretized into $T=200$ time points.
In our simulations, we add i.i.d.\ Gaussian noise with standard deviation equal to 1\% of the maximum signal amplitude to all recordings.
Consequently, $\Gammabf_{\varepsilonbf}$ is an identity matrix rescaled by the noise variance.
For each experiment we generate $K=1000$ random speckles using a discretized variant of a speckle model from \cite{dainty2013}. 
From this model, we also compute the speckle covariance matrix $\Gammabf_{\ebf}$.
We use speckles of three different sizes in the experiments to demonstrate how finer speckles lead to resolution improvement. 
Examples of speckles of each size are given in Fig.\ \ref{fig:speckle}.
We choose the parameters in \eqref{eq:photoacoustics-model} to correspond to an experiment in water.

\begin{figure}[ht!]
	\centering  
	\includegraphics[width=1\columnwidth]{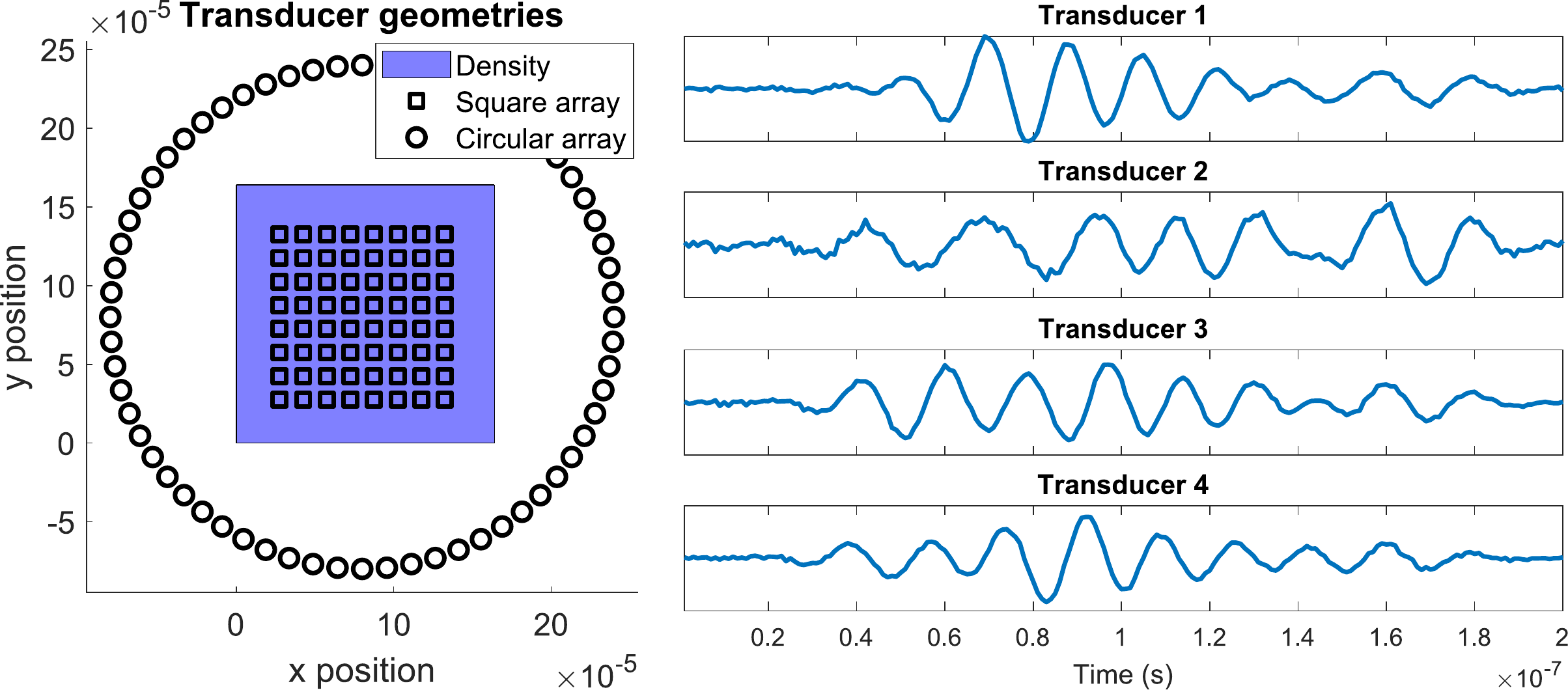}
	\caption{Left: The two transducer geometries. 
		Right: Examples of recordings from four different transducers in the circular geometry.}
	\label{fig:transducer-geometries}
\end{figure}

\begin{figure}[ht!]
	\centering  
	\includegraphics[width=1\columnwidth]{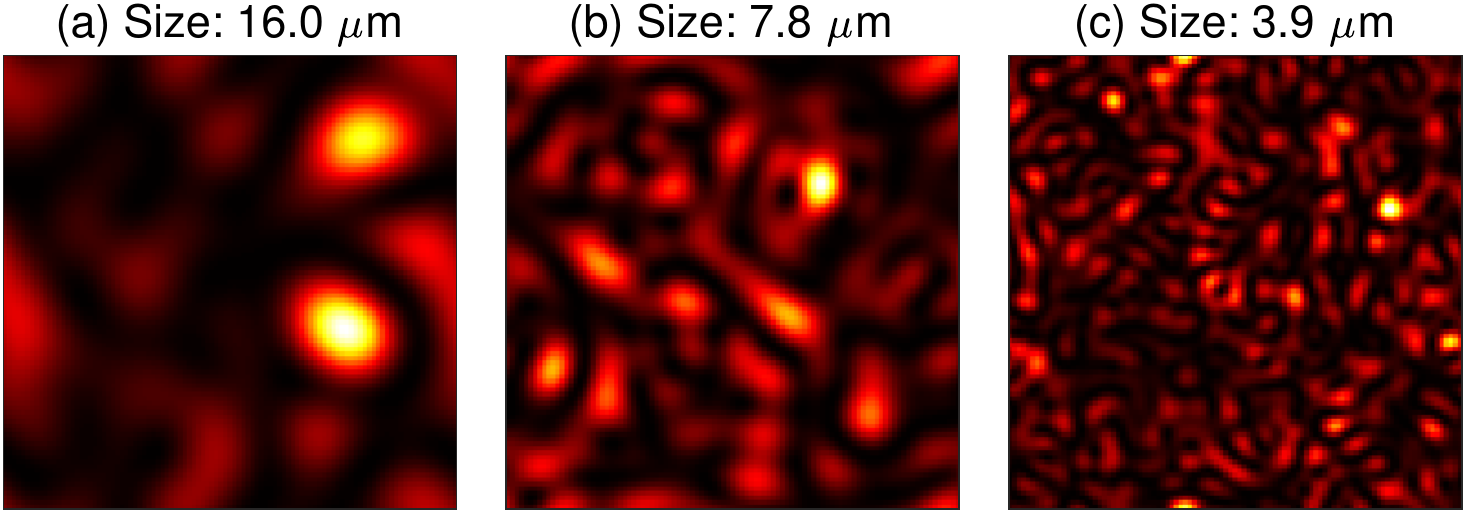}
	\caption{Examples of speckle patterns of different size.}
	\label{fig:speckle}
\end{figure}

Figs.~\ref{fig:square-array} and \ref{fig:circ-array} show the experiment results for the square and circular transducer array geometries, respectively.
Since both $\hat{\rhobf}_1$ and the time reversal solution are computed from the mean signal, they are not impacted by the speckle size.
In both figures, subplots (b) and (c) show the reconstructions by time reversal in k-Wave and via the first order method in \eqref{eq:rho-hat-1}, respectively.
Subplots (d)--(f) show how the resolution for reconstruction via Alg.\ \ref{alg:recovery} improves as the speckle size is reduced. 
The speckle sizes are those specified in Fig.\ \ref{fig:speckle}.
These experiments indicate that combining random speckle illumination and second order statistics allows us to outperform the first order methods.
In particular, finer speckles allow us to recover finer details.
We found that using speckles finer than those shown in Fig.\ \ref{fig:speckle} (c) did not lead to any further improvement in resolution.  

\begin{figure}[ht!]
	\centering  
	\includegraphics[width=1\columnwidth]{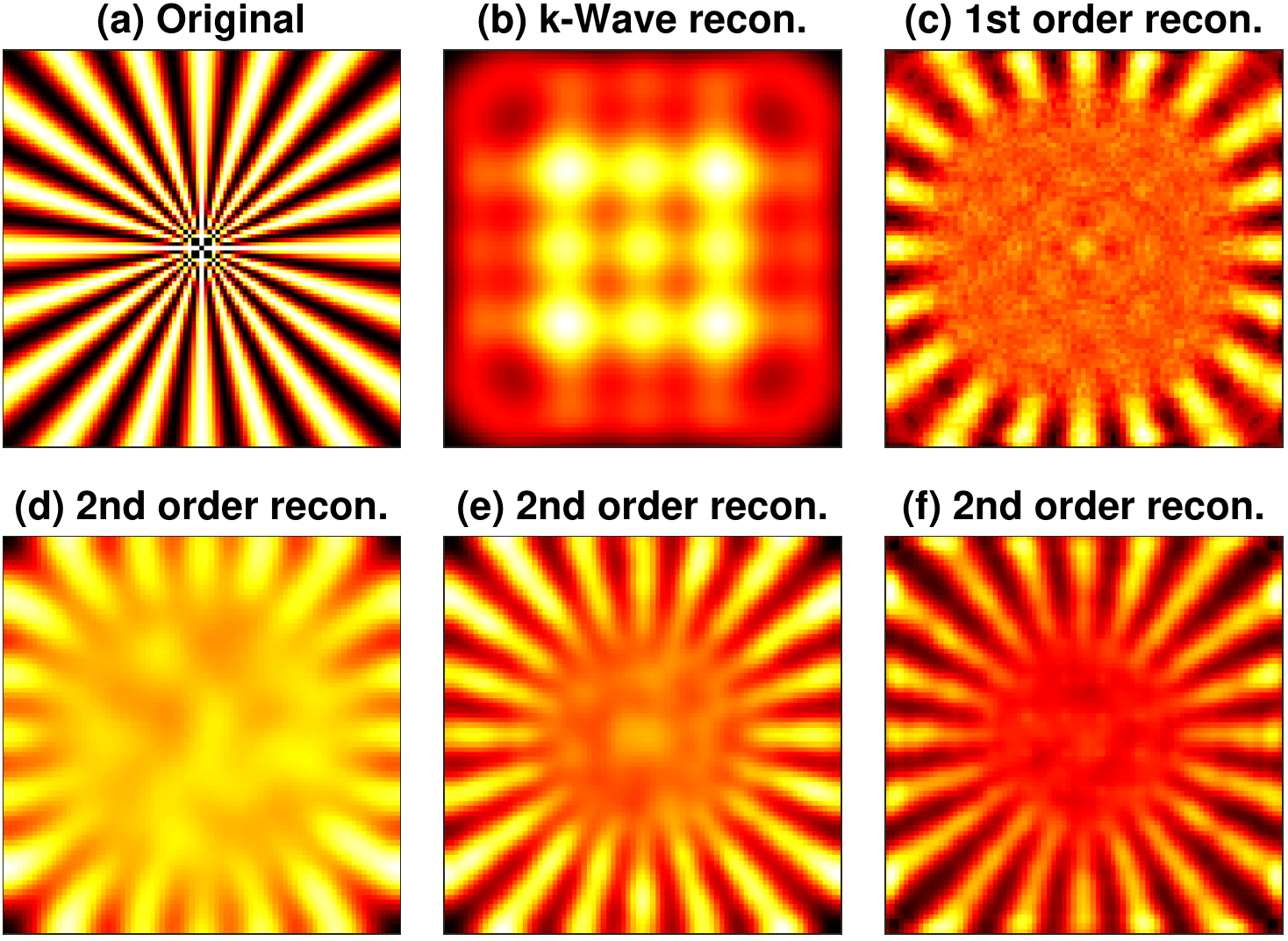}
	\caption{Reconstruction with the \emph{square} transducer array. 
		(a) Original object. 
		(b) Reconstruction using time reversal in k-Wave.
		(c) Object reconstructed via \eqref{eq:rho-hat-1}. 
		(d)--(f) Object reconstructed using Alg.\ \ref{alg:recovery} for the different speckle sizes illustrated in Fig.\ \ref{fig:speckle}.}
	\label{fig:square-array}
\end{figure}

\begin{figure}[ht!]
	\centering  
	\includegraphics[width=1\columnwidth]{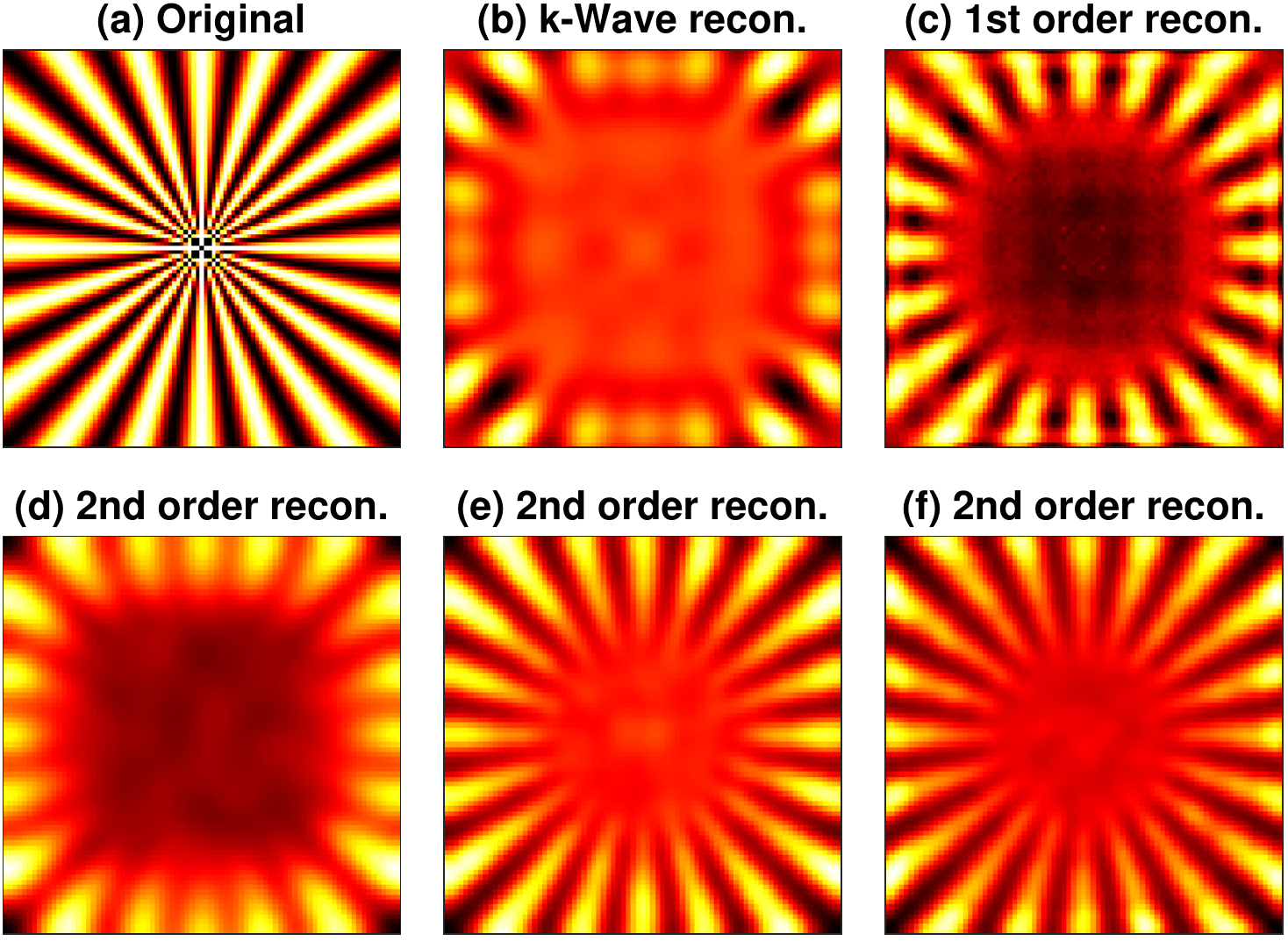}
	\caption{Reconstruction with the \emph{circular} transducer array. 
		The different subplot descriptions are the same as in Fig.\ \ref{fig:square-array}.}
	\label{fig:circ-array}
	\vspace{-1em}
\end{figure}

Our algorithm also works well in the original microscopy setting considered in \cite{idier2018}, in which case $\Abf$ just represents convolution with the PSF $h$ in \eqref{eq:idier-model}.
Indeed, we are able to achieve the same results as in \cite{idier2018} by using our algorithm at a fraction of the cost.
Due to space constraints, we do not include those results here.

\section{Conclusion} \label{sec:conclusion}

We have shown in experiments that the ideas by Idier et al.\ \cite{idier2018} in the microscopy setting can be extended to the more general PAT setting.
We also proposed a simple algorithm for computing the object which is much faster to run and easier to implement than the iterative method in \cite{idier2018}.
Despite the speedup achieved by our algorithm, it still remains quite expensive at a cost of $O(N^3)$ where $N$ is the number of pixels.
Another factor that will impact the performance of the method is how well we are able to model or estimate the true speckle covariance $\Gammabf_{\ebf}$.
Addressing these issues are an interesting direction for future research.
Other interesting directions include trying to reconstruct three-dimensional objects, and modifying Alg.~\ref{alg:recovery} to leverage prior knowledge about object sparsity.
It may also be possible to combine least squares solves in Alg.~\ref{alg:recovery} and use e.g.\ LSQR \cite{paige1982} to achieve further speedups.

\bibliographystyle{IEEEtran}
\bibliography{library}

\end{sloppy}
\end{document}